\documentclass[a4paper]{manuscript}

\usepackage[T1]{fontenc}

\title{Chaos and noise in evolutionary game dynamics}
\shorttitle{Chaos and noise}

\author[1,2,*]{María Alejandra Ramírez}
\author[3]{George Datseris}
\author[1]{Arne Traulsen}

\affiliation[1]{Max Planck Institute for Evolutionary Biology, Plön 24306, Germany}
\affiliation[2]{Max Planck Institute for Mathematics in the Sciences, Leipzig 04103, Germany}
\affiliation[3]{Department of Mathematics and Statistics, University of Exeter, EX4 4QF Exeter, United Kingdom}

\affiliation[*]{\texttt{ramirez@evolbio.mpg.de}}

\abstract{Evolutionary game theory has traditionally employed deterministic models to describe population dynamics. These models, due to their inherent nonlinearities, can exhibit deterministic chaos, where population fluctuations follow complex, aperiodic patterns.
Recently, the focus has shifted towards stochastic models, quantifying fixation probabilities and analysing systems with constants of motion. 
Yet, the role of stochastic effects in systems with chaotic dynamics remains largely unexplored within evolutionary game theory.
This study addresses how demographic noise -- arising from probabilistic birth and death events -- impacts chaotic dynamics in finite populations. We show that despite stochasticity, large populations retain a signature of chaotic dynamics, as evidenced by comparing a chaotic deterministic system with its stochastic counterpart. More concretely, the strange attractor observed in the deterministic model is qualitatively recovered in the stochastic model, where the term deterministic chaos loses its meaning.
We employ tools from nonlinear dynamics to quantify how the population size influences the dynamics. We observe that for small populations, stochasticity dominates, overshadowing deterministic selection effects. However, as population size increases, the dynamics increasingly reflect the underlying chaotic structure. 
This resilience to demographic noise can be essential for maintaining diversity in populations, even in non-equilibrium dynamics. Overall, our results broaden our understanding of population dynamics, and revisit the boundaries between chaos and noise, showing how they maintain structure when considering finite populations in systems that are chaotic in the deterministic limit.\newline}

\keywords{\hspace{4.5mm}Evolutionary game theory | Population dynamics | Chaotic dynamics | Demographic noise}

\begin{document}

\maketitle

\begin{refsection}

The question ``How does a population of interacting individuals change over time?'' has been central to multiple fields, including ecology, sociology, economics and computer science \cite{broom:book:2013, mcnamara:book:2020, raihani:book:2021, hoffman:book:2022}. 
To address this fundamental question, evolutionary game theory was developed -- a mathematical framework that applies game theory principles to model how interactions among different types of individuals drive changes within a population \cite{vonneumann:book:2007, maynard-smith:Nature:1973, hofbauer:book:1998}. In this framework, interactions are described by a payoff matrix $\pi_{ij}$, as in classical game theory.
The population's composition, represented by the relative abundances of the $d$ different types ($x_1, ..., x_d$), changes over time in response to selection. This dynamic process is modelled through frequency-dependent fitness, expressed as $f_i = \sum_{j=1}^{d} \pi_{ij}x_j$, where the success of each type $i$ depends on how common other types are within the population.
Among the various models used in evolutionary game theory, the replicator equation has emerged as the most widely used due to its simplicity and analytical tractability \cite{taylor:MB:1978,zeeman:LN:1980,schuster:JTB:1983,hofbauer:book:1998,tarnita:biorxiv:2024}. 
Notably, this model can exhibit deterministic chaos — aperiodic dynamics that exhibit sensitive dependence on initial conditions, despite the system being governed by deterministic rules \cite{skyrms:JLLI:1992,sato:PNAS:2002}. 

Understanding and describing such dynamical phenomenon has proven to be a significant challenge.
Since the groundbreaking work on chaotic dynamics in the early 1960s by Ellen Fetter, Margaret Hamilton, and Edward Lorenz, the scientific community has extensively explored how complexity and nonlinear interactions give rise to disorder \cite{lorenz:JAS:1963, schuster:book:1995,hilborn:book:2000, ott:book:2002}. In this context, chaos has emerged as a powerful theoretical tool, revealing how simple nonlinear, low-dimensional models can lead to aperiodic and seemingly unpredictable population fluctuations \cite{may:Nature:1976}. 

In the past two decades, interest has grown in stochastic models for evolutionary game dynamics, particularly for capturing the inherent randomness of real-world populations \cite{nowak:Nature:2004, traulsen:PRE:2006b, traulsen:PRE:2012}. However, studies of chaos in evolutionary game dynamics has remained limited to the case of infinite, well-mixed populations \cite{sato:PNAS:2002,galla:PNAS:2013,sanders:SciRep:2018}.
Chaotic dynamics in finite populations remains largely unexplored, mainly due to the stochasticity introduced by the population finiteness. This stochasticity, known as demographic noise, stems from the inherent unpredictability of birth and deaths within a population. In this manuscript, we investigate how demographic noise interacts with the complexity of chaotic dynamics.

To explore this interplay, we employ the pairwise comparison process, a model that enables a comprehensive analysis of how population size and selection intensity influence the system's dynamics \cite{blume:GEB:1993,szabo:PRE:1998,traulsen:bookchapter:2009a}. To quantify the key features of the dynamics, we apply numerical measures from nonlinear time series analysis, providing readily comparable metrics across the parameter regime \cite{kantz:book:2003, datseris:book:2022}.

We investigate how demographic noise affects chaotic dynamics and confirm that in small populations, stochasticity overshadows selection, dominating the system’s dynamics. However, as population size increases, the dynamics increasingly reflect the underlying deterministic system, allowing for chaotic dynamics to emerge. Remarkably, the chaotic attractor persists despite the constant perturbations from demographic noise. This is significant because chaotic systems are highly sensitive to small changes, yet here the dynamics remain robust, closely following the attractor of the deterministic system. Consequently, we find that in finite, but sufficiently large populations, the boundary between chaos and noise blurs, suggesting that signatures of chaos can still appear in stochastic models.

\section*{\large Model}
In this manuscript, we analyse the interplay between chaotic dynamics and demographic noise using evolutionary game theory. This framework has two key components: the payoff matrix and the update mechanism. The payoff  matrix represents the interactions between the different types in the population, while the update mechanism describes how the population composition changes over time according to a given rule.

\vspace{2mm}
\textbf{\small Payoff matrix.}
To analyse low-dimensional chaos in the context of evolutionary game theory, we employ the ACT-Skyrms payoff matrix. 
The payoff matrix originates from the work of Arneodo, Coullet and Tresser (ACT), who constructed an interaction matrix capable of displaying complex dynamics in the framework of Lokta-Volterra systems \cite{arneodo:PLA:1980, arneodo:JMB:1982}. They employed Shilnikov's theorem to construct a three-dimensional system displaying a strange attractor and provide numerical evidence to support its existence \cite{shilnikov:PUAS:1965} (see SI Appendix). The resulting Shilnikov-type attractor is a specific kind of chaotic attractor associated with a homoclinic orbit that returns to a saddle-focus equilibrium point. These equilibrium points have both stable and unstable directions, where trajectories can be attracted to them in some directions while being repelled in others. In a Shilnikov attractor, trajectories are initially repelled from the equilibrium point but eventually return along a homoclinic orbit. As they loop around the saddle focus, they create complex dynamics known as spiral or Shilnikov chaos \cite{guckenheimer:book:2013, hirsch:book:2013, gonchenko:Chaos:2022}. 

\begin{figure}[t!]
    \centering
    \includegraphics[width=1.0\linewidth]{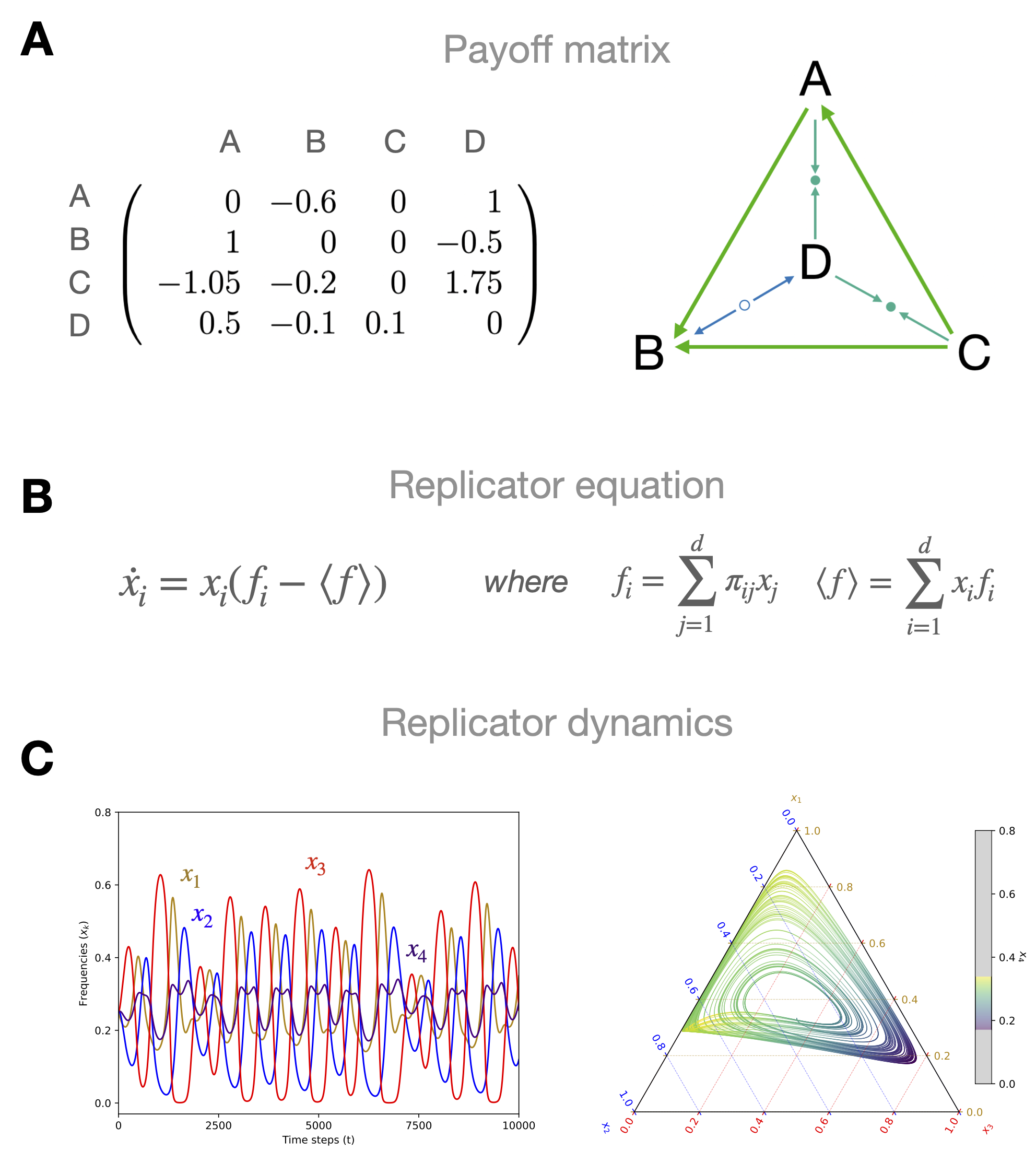}
    \caption{The replicator equation for the ACT-Skyrms payoff matrix exhibits aperiodic dynamics characterised by a Shilnikov strange attractor. \textbf{A.} ACT-Skyrms payoff matrix and the corresponding relations between different strategies. \textbf{B.} The replicator equation is the standard update mechanism to model the evolution of an infinitely large, well-mixed population. It states that the change in abundance of a given strategy $x_i$ is proportional to the difference between the individual fitness $f_i$ and the average fitness of the population $\langle f \rangle$. \textbf{C.} The replicator equation for the ACT-Skyrms payoff matrix exhibits aperiodic dynamics characterised by a Shilnikov strange attractor. The trajectories display 10,000 time steps starting from the initial condition $\vec{x}=(0.25,0.25,0.25,0.25)$.}
    \label{fig:payoffmatrix_repdyn}
\end{figure}

Based on the ACT construction, Skyrms later formulated the game dynamical counterpart of the ACT interaction matrix, creating the so-called ACT-Skyrms payoff matrix (Fig. \ref{fig:payoffmatrix_repdyn}A) \cite{skyrms:JLLI:1992}. This formulation is based on the mapping between a Lokta-Volterra system with $n$ species and an evolutionary game with $n+1$ types or strategies \cite{hofbauer:book:1998}. The ACT-Skyrms payoff matrix describes the following relations between the different strategies: A dominates C, B dominates A and C, B and D are bistable, A and D coexist, as well as C and D (Fig. \ref{fig:payoffmatrix_repdyn}A). In terms of classical game theory, the only dominated strategy is A.

In his paper, Skyrms provided numerical evidence for the existence of the Shilnikov-type attractor within the framework of evolutionary game theory using the replicator equation, the standard update mechanism to model the evolution of an infinitely large, well-mixed population. As shown in figure \ref{fig:payoffmatrix_repdyn}B, the replicator equation explains how the relative abundance -or frequency- of a strategy, $x_i$, changes over time, 
\begin{equation}
      \label{eq:RepDyn}
      \dot x_i = x_i \left(f_i - \langle f \rangle \right).
\end{equation}
A strategy's abudance increases if its individual fitness, $f_i$, is higher than the average fitness of the population $\langle f \rangle$. Conversely, if its fitness is lower than average, its abundance decreases \cite{taylor:MB:1978,zeeman:LN:1980,schuster:JTB:1983}.
The focus of the replicator dynamics on relative abundances $\vec{x}=(x_1, ..., x_d)$, reduces the system's effective dimension from $d$ to $d-1$, due to the normalisation condition
\begin{equation}
    \label{eq:normalisation}
    \sum^{d}_{i=1} x_i = 1.
\end{equation}
For example, the 4x4 ACT-Skyrms payoff matrix leads to a set of 4 coupled differential equations that define a 3-dimensional continuous system.

Figure \ref{fig:payoffmatrix_repdyn}C shows that the replicator equation for the ACT-Skyrms payoff matrix exhibits aperiodic dynamics, characterised by a Shilnikov strange attractor. Therefore, we employ the ACT-Skyrms payoff matrix to analyse low-dimensional chaos in the framework of evolutionary game theory.

\vspace{2mm}
\textbf{Update mechanism - Pairwise comparison process (PCP).}
The replicator equation describes an infinitely large population. 
In reality, however, populations are finite and, thus, subject to demographic noise, which is a source of stochasticity stemming from the discrete and random nature of birth and death events in populations \cite{nowak:Nature:2004,ewens:book:2004,traulsen:PRL:2005,constable:PNAS:2016}. To include demographic noise, we use the pairwise comparison process (PCP) to model the evolution of populations \cite{blume:GEB:1993,szabo:PRE:1998,traulsen:bookchapter:2009a}. This formulation extends the replicator equation, allowing a free exploration of the selection intensity and facilitating the analysis of evolutionary dynamics within the context of stochastic processes \cite{gardiner:book:2004}. Therefore, PCP's primary advantage is to provide a comprehensive framework for analysing any arbitrary evolutionary game in finite populations across all selection intensities \cite{wu:NJP:2015}. In the PCP formulation, a pair of individuals, a focal individual ($j$) and a role model ($i$), are randomly selected from the population. Subsequently, the focal individual adopts the role model's strategy according to a replacement probability defined in terms of the strategies' fitness difference.

\begin{equation}
    \label{eq:adopt_prob}
    p = \frac{1}{1+e^{-\beta(f_i - f_j)}}
\end{equation}
$f_i$ is the role model's fitness and $f_j$ is the focal individual's fitness.
The selection intensity, $\beta>0$, determines how strongly the game influences the dynamics of a population with constant size $N$, leading to different regimes for the dynamics \cite{blume:GEB:1993, hauert:AJP:2005}.
If $\beta=0$, the dynamics are governed by neutral drift. 
If $\beta \ll 1$, the adoption probability is approximately linear with respect to fitness differences, so the replicator dynamics is recovered. 
Finally, if ${\beta \gg 1}$, the dynamics is dominated by the evolutionary game described by the payoff matrix.
    
To analyse the interplay between chaos and demographic noise, we use PCP as an individual-based model, which allows both numerical and analytical analysis. This is possible because individual-based models are formulated as stochastic Markov processes, explicitly accounting for the inherent discreteness and randomness of the system \cite{black:TREE:2012}.
Specifically, we describe the evolutionary changes of a population with constant size $N$ as a birth-death process, which results in the following transition probabilities 
\begin{equation*}
    \label{eq:TransProb_PCP}
    \small
    T_{ji}(\vec{x})  = x_{i}x_{j}\frac{1}{1+e^{-\beta(f_i - f_j)}} \hspace{4mm} T_{ij}(\vec{x})  = x_{i}x_{j}\frac{1}{1+e^{+\beta(f_i - f_j)}}.
\end{equation*}
    
$T_{ji}(\vec{x}) $ is the probability that the abundance of type $i$ individuals changes from $x_i$ to $x_i + \frac{1}{N}$, while the abundance of type $j$ changes from $x_j$ to $x_j - \frac{1}{N}$. Similarly, $T_{ij}(\vec{x})$ represents the case where the abundance of $j$ increases.
    
Chaotic dynamics is defined in the realm of deterministic systems. Thus, we require a deterministic description of the pairwise comparison process (PCP). 
For this purpose, we derive the population-level model of the PCP for $d$ different types performing a Kramers-Royal expansion of the master equation, the fundamental equation governing the dynamics of the stochastic process \cite{kampen:book:1997,gardiner:book:2004,black:TREE:2012}.
This procedure yields a Fokker-Planck equation that is equivalent to a stochastic differential equation. According to Itô calculus, the process can be approximated by a Langevin equation composed of a deterministic drift term $a_i(\vec{x})$ and a diffusion matrix \cite{helbing:TD:1996,traulsen:PRE:2006b,traulsen:PRE:2012}. Thus, the process can be written as a stochastic differential equation, where in the limit of very large population sizes ($N \rightarrow \infty$) the diffusion matrix term vanishes with $~1/\sqrt{N}$. Consequently, only the drift term remains, $a_i(\vec{x})$, and we recover a set of deterministic differential equations, 
\begin{equation*}
    \label{eq:LangEq_ok_PCP}
    \dot{x_i} = a_{i}(\vec{x}) = \sum_{j=1}^{d} \left[T_{ji}(\vec{x}) - T_{ij}(\vec{x})\right].
\end{equation*}
After simplifying the equation, we obtain the population-level model for the generalised pairwise comparison process
\begin{equation}
    \label{eq:Det_PCP}
    \dot{x_i} = \sum_{j=1}^{d} x_ix_j\tanh\left[\frac{\beta}{2}(f_i(\vec{x}) - f_j(\vec{x}))\right] \hspace{5mm} i=1,..,d.
\end{equation}
Under weak selection, \(\beta \ll 1\), the approximation $\tanh(x) \approx x$ recovers the replicator equation (\eqref{eq:RepDyn}), with an additional time rescaling factor of $\beta/2$.

\section*{Results}

As indicated by numerical evidence and various quantitative measures, chaotic dynamics emerge in the low selection intensity regime.
In this regime, the deterministic population-level model exhibits confined aperiodic dynamics that depend sensitively on initial conditions. When demographic noise is introduced in the stochastic individual-based model, the trajectories continue to reflect the chaotic underlying structure, particularly for large population sizes. Conversely, in smaller populations, demographic noise overshadows the selection effects responsible for the underlying dynamics.

\subsection*{\normalsize Deterministic description}

\begin{figure}
      \centering
      \includegraphics[width=0.9\linewidth]{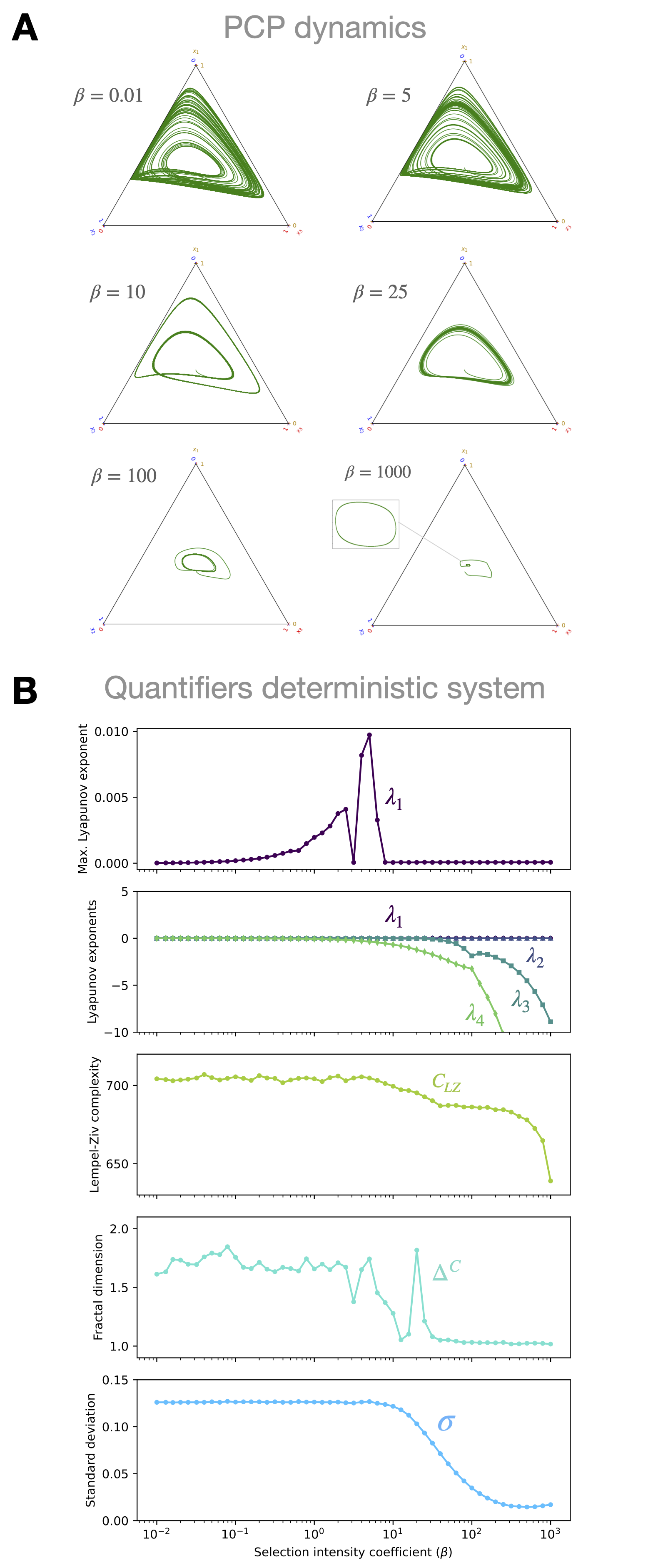}
      \caption{Deterministic chaos arises under low selection intensity. \textbf{A.} Ternary plots of the deterministic PCP dynamics for different selection intensity values ($\beta$). For low selection strength, the trajectories display a strange attractor, characterised by its fractal structure.
      For higher selection intensities, the system exhibits periodic dynamics, where the attractor gradually contracts in the state space towards the equilibrium point $x^*$. The trajectories display $1\times 10^4$ effective time steps starting from the initial condition $\vec{x(0)} = (0.25,0.25,0.25,0.25)$.
      \textbf{B.} Quantification of the chaotic dynamics using various numerical measures. A critical value, $\beta^* \approx 7$, is identified, which distinguishes chaotic from non-chaotic behavior. For $\beta < \beta^*$, the system exhibits chaotic behavior characterised by a positive maximum \textbf{Lyapunov exponent}, $\lambda_1$, indicating sensitive dependence on initial conditions. Additionally, the system displays a higher \textbf{Lempel-Ziv complexity}, $C_{LZ}$, suggesting a lack of periodicity. In this chaotic regime, the system possesses a strange attractor with a non-integer \textbf{fractal dimension}, $\Delta^C$, that spans a larger region of the state space, as indicated by a higher \textbf{standard deviation}, $\sigma$.
      }
      \label{fig:PCP_deterministic}
\end{figure}

\textbf{\small Chaotic dynamics arise under low selection intensity.}
To analyse low-dimensional chaos in the framework of evolutionary game theory, we employ a population-level model, namely the deterministic pairwise comparison process (\eqref{eq:Det_PCP}) with the ACT-Skyrms payoff matrix describing the interactions (Fig. \ref{fig:payoffmatrix_repdyn}A). We find that the dynamics of the trajectories differ depending on the selection intensity coefficient ($\beta$). Overall, a low selection strength introduces the required structural instability that allows for chaotic dynamics to emerge. Figure \ref{fig:PCP_deterministic}A shows that for the low selection intensity regime the strange attractor reported by Skyrms is recovered \cite{skyrms:JLLI:1992}. In fact, the PCP can be approximated to the replicator dynamics in the limit small values of $\beta$, with an additional time scaling (see SI Appendix).

For higher selection intensities, the region to which the trajectories are confined becomes smaller, and the dynamics become increasingly periodic. This occurs because the hyperbolic tangent function in \eqref{eq:Det_PCP} saturates at large values of $\beta$, causing the previously smooth function to act like a step function. In particular, the trajectory contracts to the unique interior equilibrium point $x^* = (0.274, 0.215, 0.241, 0.269)$. Interestingly, for all $\beta$ values the trajectory average also coincides with $x^*$. This can be explained by the fact that $x^*$ is the unstable fixed point around which the trajectories of the Shilnikov attractor loop.

After providing first evidence of the strange attractor in figure \ref{fig:PCP_deterministic}A, we proceed to quantify the main features of the system's dynamics in figure \ref{fig:PCP_deterministic}B. For this purpose, we use several numerical measures that effectively describe the observed dynamics. Specifically, we quantify the dynamics by considering characteristics related to the instability (Lyapunov exponents), periodicity (Lempel-Ziv complexity measure and Fourier spectrum), and geometry (Fractal dimension and Standard devation) of the trajectories.

\vspace{2mm}
\textbf{\small The system exhibits sensitive dependence on initial conditions.}
The Lyapunov exponents quantify the stability with respect to infinitesimal perturbations. The exponents are key indicators of chaotic dynamics because they quantify the sensitivity to initial conditions. In chaotic systems, even infinitesimal perturbations to the initial conditions cause trajectories to diverge exponentially over time. After time $t$, the distance between the original and perturbed trajectories can be approximated as $\delta(t) \approx \delta_0 e^{\lambda_1 t}$. Here, $\lambda_1$, the largest Lyapunov exponent, measures the rate of this exponential divergence of nearby trajectories. A positive $\lambda_1$ indicates chaos, reflecting the system's sensitive dependence on initial conditions \cite{hilborn:book:2000,ott:book:2002,datseris:book:2022}.

\begin{table}[h!]
    \centering
     \scalebox{0.8}{
     \begin{tabular}{|c | c c c c|}
     \hline
     $\beta$ & $\lambda_1$ & $\lambda_2$ & $\lambda_3$ & $\lambda_4$ \\ [0.5ex]
     \hline
     0.01 & $1.9 \times 10{-5}$ & $6.9 \times 10{-7}$ & $-2.3 \times 10{-7}$ & $-7.2 \times 10{-4}$ \\ 
     0.1 & $1.9 \times 10{-4}$ & $5.1 \times 10{-6}$ & $-2.2 \times 10{-6}$ & $-7.2 \times 10{-3}$ \\
     1.0 & $1.9 \times 10{-3}$ & $2.2 \times 10{-5}$ & $-1.3 \times 10{-5}$ & $-7.2 \times 10{-2}$ \\ \arrayrulecolor{gray}\hline
     10.0 & $1.6 \times 10{-5}$ & $-1.7 \times 10{-5}$ & $-1.8 \times 10{-3}$ & $-6.8 \times 10{-1}$ \\
     100.0 & $5.2 \times 10{-6}$ & $-5.2 \times 10{-6}$ & $-1.9$ & $-3.2$ \\
     1000.0 & $7.0 \times 10{-6}$ & $-4.3 \times 10{-6}$ & $-8.9$ & $-35.1$ \\ [1ex] 
     \hline
     \end{tabular}}
     \caption{\normalfont{Lyapunov exponents for different $\beta$ values. We numerically identify a critical value, $\beta^* $, which separates chaotic from non-chaotic dynamics in the system. For $\beta < \beta^*$, the system exhibits chaotic dynamics, characterised by a positive maximum Lyapunov exponent $\lambda_1$.
     Moreover, the system exhibits two near-zero exponents: one corresponds to the zero exponent associated with the flow of a continuous dynamical system, and the other results from an effective reduction in the system's dimensionality. The near-zero exponents are identified by being at least two orders of magnitude smaller than the other exponents.
     Additionally, for $\beta < \beta^*$, the exponents scale proportionally with $\beta$, which can be explained by the inverse proportionality between the system's time scaling and $\beta$.}}

    \label{table:LE}
\end{table}

\vspace{-2mm}

Figure \ref{fig:PCP_deterministic}B and table \ref{table:LE} present a critical value, $\beta^*$, which separates chaotic from non-chaotic dynamics in the system. Numerical analysis indicates that this critical value is around $\beta^* \approx 7$. For $\beta < \beta^*$, the system exhibits chaotic dynamics, indicated by a positive largest Lyapunov exponent $\lambda_1$.

In general, the system is described by four Lyapunov exponents, in our measurements two of them are near-zero exponents. This may be attributed to the fact that, despite the system being defined by four differential equations, the normalisation condition described by \eqref{eq:normalisation} effectively reduces the dimension from $d$ to $d-1$. Alternatively, it could be related to the nature of the attractor. In systems with Shilnikov attractors, trajectories often loop around a lower-dimensional manifold, reducing the effective dimensionality \cite{gonchenko:IJBC:2005,grines:Chaos:2022}. As a result, one near-zero exponent is associated with the direction of the flow in the continuous system, while the other reflects the reduced dimensionality of the dynamics. In our system, these two near-zero exponents are identified by being at least two orders of magnitude smaller than the other exponents.

More precisely, for $\beta < \beta^*$, where chaotic dynamics are observed, the Lyapunov spectrum consists of one positive exponent ($\lambda_1$), two near-zero exponents ($\lambda_2$, $\lambda_3$), and one negative exponent ($\lambda_4$). For ${\beta > \beta^*}$, there is a topological change and the system transitions to non-chaotic dynamics, displaying two near-zero exponents (${\lambda_1,\lambda_2}$) and two negative exponents (${\lambda_3, \lambda_4}$). As $\beta$ increases, the negative exponents become increasingly negative, consistent with the increased contraction of the attractor. Additionally, for $\beta < \beta^*$, the exponents scale proportionally with $\beta$, which can be explained by the inverse proportionality between the system's time scaling and $\beta$ (since the exponents have units of inverse time; see SI Appendix). Thus, the predictability horizon defined as $1/\lambda_1$ remains constant for the chaotic regime \cite{vulpiani:book:2010, cencini:JPA:2013}. 
Finally, we measure the sum of the exponents and confirm that it is negative, as expected for dissipative systems.

\vspace{2mm}
\textbf{\small The trajectories display aperiodic dynamics.}
Chaotic dynamics are characterised by aperiodic dynamics. To identify this type of dynamics, we use the Lempel-Ziv complexity ($C_{LZ}$). This quantifier measures the repetitiveness of a finite sequence, making it a useful tool for quantifying the periodicity of a time series. The algorithm counts the number of distinct patterns that comprise a data set. A lower $C_{LZ}$ value indicates a less complex sequence or a more repetitive data set. As shown in Figure \ref{fig:PCP_deterministic}B, as $\beta$ increases, the Lempel-Ziv complexity decreases. This result aligns with the observation that, for high selection intensity, trajectories become more periodic and require fewer unique patterns due to their reduced complexity. We obtain the same results through a periodicity analysis using the Fourier spectrum (see SI Appendix), which further confirms this behavior.

\vspace{2mm}
\textbf{\small The dynamics are characterised by a strange attractor.}
Chaotic systems are characterised by ``strange attractors'', which are dynamical attractors that exhibit infinitely repeating fine structures. The fractal dimension ($\Delta^{c}$) captures this key geometric feature because it is an invariant measure that reflects the system’s dimensionality by accounting for the self-similar structure of the attractor \cite{grassberger:PRL:1983, addison:book:1997}.
Figure \ref{fig:PCP_deterministic} shows that, under low selection intensity, the attractor exhibits a fractal dimension of around $1.7$. This value corresponds to a curve with a repeating fine structure, indicating that the attractor is topologically situated between a curve and a surface. In other words, the system exhibits a strange attractor, where a curve winds repeatedly through the state space. As the selection intensity ($\beta$) increases, the fractal dimension decreases, reflecting the progression toward periodic dynamics. Eventually, when $\beta > \beta^*$, the fractal dimension reaches a value of $1.0$, which is the topological dimension of a curve. This agrees with the emergence of a limit cycle in the high selection regime.

\vspace{2mm}
\textbf{\small The attractor contracts for high selection intensity.}
In this work, the standard deviation ($\sigma$) is used to quantify the attractor's contraction as the selection intensity ($\beta$) increases. Figure \ref{fig:PCP_deterministic} shows that, under low selection intensity, the attractor covers a greater region of the state space, indicated by a higher standard deviation. As $\beta$ increases, the attractor contracts towards the unstable fixed point, $x^*$, which is reflected by a smooth decrease in the average variability of the trajectories.

\begin{figure*}[t!]
    \centering
    \includegraphics[width=1.0\linewidth]{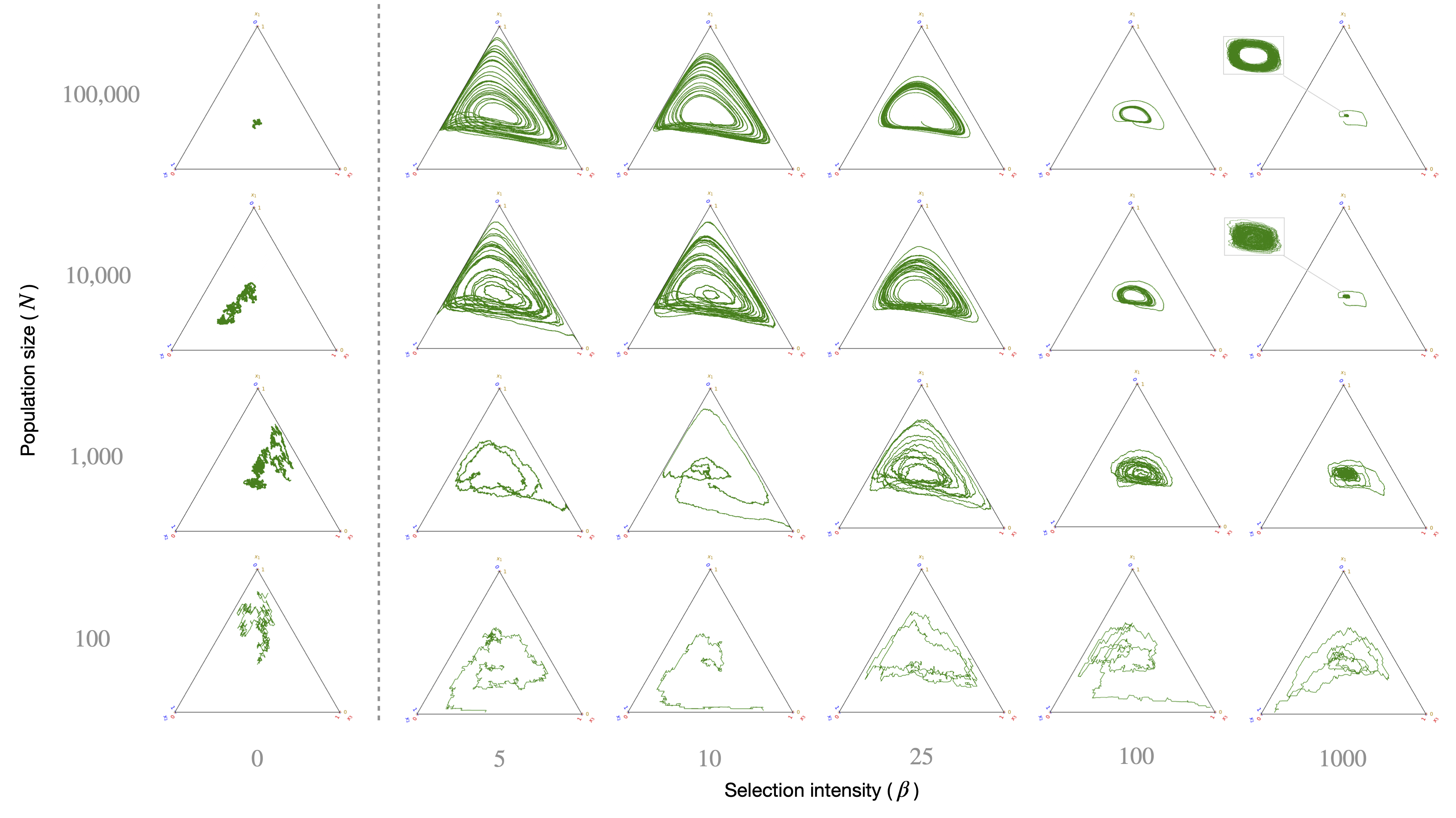}
    \caption{The underlying chaotic structure is reflected in the stochastic dynamics for large populations. The stochastic trajectories show that for small populations demographic noise largely dominates the dynamics. Specifically, demographic noise overshadows selection effects. In contrast, for large population sizes the dynamics approximate the deterministic limit because the step size becomes increasingly small, $\Delta \vec{x} \rightarrow 0$. However, the transition between strange attractor and limit cycle in the presence of demographic noise is less distinct than in the deterministic case.
    Finally, regardless of the population size the system's dynamics are primarily governed by the underlying attractor. All the trajectories start from the same initial condition $\vec{x(0)} = (0.25,0.25,0.25,0.25)$ and display an equivalent number of time steps according to the scaling described in the SI Appendix.}
    \label{fig:ternary_sto}
\end{figure*}

\subsection*{\normalsize Stochastic description}

\textbf{\small The signatures of chaos emerge for large population sizes.}
Our stochastic description is based on the pairwise comparison process (PCP) as an individual-based model, where the basic entities are individuals that can adopt different  strategies according to the replacement probability in \eqref{eq:adopt_prob}. This approach allows us to freely explore how the selection intensity ($\beta$) and the population size ($N$) affect the population's composition dynamics. 

We observe that the selection intensity ($\beta$) plays a key role in determining the dynamics of the system, as shown in Figure \ref{fig:ternary_sto}. When $\beta=0$, the trajectories follow a random walk, indicating the absence of selection. However, once selection is introduced, the dynamics shift significantly. 
At low selection intensity, stochasticity is introduced into the replacement process via a smooth replacement probability, leading to the emergence of structural instability, which becomes a key driver of chaotic dynamics. The increased randomness in selection allows the system to explore a wider range of states, amplifying sensitivity to initial conditions and promoting complex behavior.
In contrast, at high selection intensity, the stochasticity in the replacement events is minimized because the replacement probability resembles a step function. Consequently, trajectories tend to stay close to the equilibrium point $x^*$, and the dynamics become more stable and predictable.

Moreover, we confirm that the population size ($N$) has an important impact on the dynamics. For small populations, the dynamics are largely dominated by stochasticity due to demographic noise.
On the contrary, for larger populations, the dynamics are strongly influenced by the underlying deterministic structure, eventually making the deterministic and stochastic trajectories nearly indistinguishable. This difference arises from the inverse proportionality between step size and population size, $\Delta \vec{x} = 1/N$, where only one birth-death event occurs in the entire population per time step (see SI Appendix). Thus, as population size increases, the effect of demographic noise diminishes, along with the influence of individual event uncertainty on the macroscopic dynamics due to smaller step sizes, $\Delta \vec{x} \rightarrow 0$.

\begin{figure}[t!]
    \centering
    \includegraphics[width=1.0\linewidth]{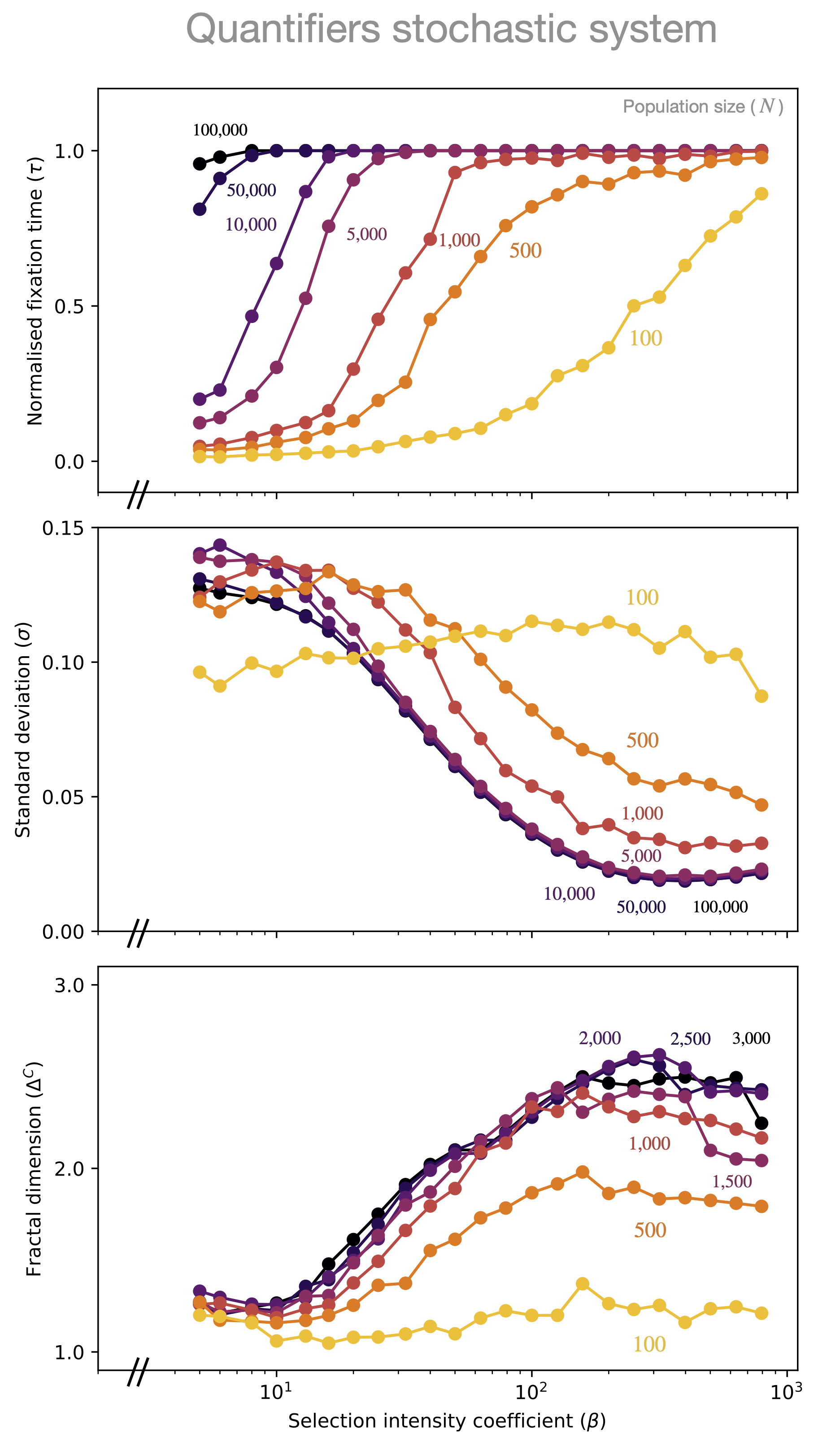}
    \caption{Quantification of the stochastic trajectories. \textbf{A.} Dynamics for smaller population sizes result in faster fixation. This occurs because the step size is larger for smaller populations, delaying the time for the trajectories to reach the boundaries signifying the extinction of one of the population's types. \textbf{B.} Demographic noise overshadows selection effects, therefore there is no attractor's contraction for smaller populations, as shown by the attractor's standard deviation. \textbf{C.} The fractal dimension increases as the selection intensity becomes larger. The perturbations generated by demographic noise prevent the trajectories from forming a well-defined limit cycle under high selection. The plot shows the average quantifier values for 100 stochastic runs.}
    \label{fig:quantifiers_sto}
\end{figure}

\vspace{2mm}
\textbf{\small The underlying attractor is robust to demographic noise.}
Figure \ref{fig:ternary_sto} illustrates that the system's dynamics are primarily governed by the underlying attractor. When trajectories start near the equilibrium point $x^*$, they are strongly influenced by the underlying topological structure causing them to loop around $x^*$.
However, as shown in figures \ref{fig:ternary_sto} and \ref{fig:quantifiers_sto}, dynamics for smaller population sizes result in faster fixation. This happens because smaller populations have a larger step size, $\Delta \vec{x}$, which increases the risk of one of the types going extinct. If extinction occurs, the system’s dimensionality is reduced, thus chaos is no longer possible for the current system. In contrast, for large population sizes, $\Delta \vec{x}$ is smaller, causing fixation time to increase with extinction being delayed.

\vspace{7mm}
\textbf{\small Strong demographic noise overshadows selection effects.}
Figure \ref{fig:ternary_sto} shows that the effect of the selection intensity ($\beta$) is more pronounced in the dynamics for larger populations. For instance, figure \ref{fig:quantifiers_sto} shows that the reduction of the attractor's standard deviation is greater for larger populations. Thus, the contraction of the attractor due to high selection is more evident when the stochastic effects do not play an important role. This suggests that for small populations the stochasticity caused by demographic noise outweighs selection effects.

The contraction of the attractor for large $\beta$ values is driven by the saturation of the Fermi function that defines the replacement probability in \eqref{eq:adopt_prob}. As $\beta$ increases, the Fermi function transitions into a step function. This effect mirrors the dynamics in the deterministic system, where the saturation of the hyperbolic tangent in \eqref{eq:Det_PCP} causes the transition from chaotic to non-chaotic dynamics. Consequently, in both the deterministic and stochastic models, high selection intensity results in the contraction of the attractor because a previously smooth function saturates to behave like a step function.

For low selection strength, the replacement probability (\eqref{eq:adopt_prob}) is approximately linear with respect to fitness differences, allowing for the replicator dynamics to be recovered. Consequently, we observe the strange attractor of the deterministic description, shown in figures \ref{fig:payoffmatrix_repdyn} and \ref{fig:PCP_deterministic}, emerging from an exclusively stochastic description.


\vspace{2mm}
\textbf{\small Stochasticity blurs the transition between strange attractors and limit cycles.}
The transition between strange attractor and limit cycle in the presence of demographic noise is less distinct than in the deterministic case (see figures \ref{fig:PCP_deterministic} and \ref{fig:ternary_sto}). This ambiguity arises because demographic noise disrupts the trajectories, preventing them from forming a well-defined limit cycle around the saddle-focus. Therefore, the stochastic trajectories looping around the equilibrium point $x^*$ have an additional layer of fine structure that is included into the dynamics by the finiteness of the population. 

For example, figure \ref{fig:ternary_sto} shows that in large populations, high selection intensity contracts the attractor, causing stochastic trajectories to cluster closely together. This creates a structure that resembles a plane rather than a curve, shaped by the microstructure introduced by stochastic effects.
As a result, in figure \ref{fig:quantifiers_sto} the fractal dimension increases with increasing $\beta$. In contrast, the dimension of the deterministic trajectories decreases as $\beta$ increases (see figure \ref{fig:PCP_deterministic}).

Overall, the fractal dimension quantifies the empirical observations from both deterministic and stochastic trajectories, and it effectively distinguishes chaos from noise. This is because the fractal dimension inherently captures the fine structure of a set. In stochastic systems, it is able to differentiate between the self-similar microstructure of noise and the fractality of the strange attractor when the self-repeating structures emerge at different spatial scales, as it is further discussed in the SI Appendix.

\section*{Discussion}
The emergence of disorder in populations composed of interacting individuals has been a long-lasting question across various fields, including ecology and economics \cite{sato:PNAS:2002, dennis:OIKOS:2003, ellner:OIKOS:2005, farmer:book:2024}. Traditionally, disorder in these systems has been attributed either to their inherent complexity or to stochastic effects. However, the scenario where complexity and stochasticity arise in the same system remains understudied, despite growing recognition that the synergy between stochastic effects and deterministic nonlinearities is crucial for accurately representing real-world phenomena \cite{higgins:Science:1997,bjornstad:Science:2001,rohani:AMNAT:2002,coulson:TEE:2004}. This challenge stems from the fact that both chaos and noise manifest as irregular temporal fluctuations, making them difficult to distinguish \cite{cencini:JPA:2013, bradley:Chaos:2015}. Nonetheless, understanding their distinct dynamical origins is crucial: chaos arises from deterministic complexity, while noise is a product of stochasticity.


In this study, we present a direct approach to disentangle the effects of chaos and noise by using evolutionary game theory to analyse how stochasticity influences chaotic population dynamics. Specifically, we apply the pairwise comparison process to compare chaotic dynamics in a deterministic population-level model with its stochastic individual-based counterpart. This approach enables us to explore the impact of demographic noise —random fluctuations from stochastic birth and death events— on chaotic dynamics. Remarkably, we find that in sufficiently large but finite populations, hallmarks of deterministic chaos persist in the purely stochastic system. The same strange attractor that drives chaos in the deterministic model re-emerges almost identical within the stochastic model, highlighting a surprising robustness of chaotic dynamics despite the presence of noise. 


We observe that the strange attractor is resilient to perturbations despite the system's inherent sensitivity to small perturbations, providing a smooth transition between chaos and noise. Even though demographic noise introduces perturbations at each time step in the system displaying non-equilibrium dynamics, the dynamics are resilient enough to maintain diversity in the population. This resilience contrasts with the noise-induced chaos that arises from the weak stability of limit cycles, where small perturbations can lead to significant changes \cite{gao:PRL:2002}. Thus, our results highlight the close relation between the effect of noise and the system's underlying topological structure.

To further explore the robustness of chaos to demographic noise, future studies could analyse different payoff matrices within the game theory framework. However, it is important to note that chaotic dynamics are more commonly observed in high-dimensional systems \cite{albers:PRE:2006,mallmin:PNAS:2024}. Additionally, our theoretical findings could be validated with experimental data from populations of varying population sizes while interacting under the same game.

In conclusion, we demonstrated that the signatures of deterministic chaos can be observed in sufficiently large populations even in the presence of inherent stochasticity. We provide a foundational proof of concept demonstrating that elements of chaos theory remain applicable to evolutionary dynamics of finite populations. For instance, our results suggest that the emergence of a strange attractor is primarily determined by the underlying topological structure of both the deterministic and the stochastic model. More broadly, we confirm that for large populations sizes, the dynamics are strongly influenced by the underlying chaotic attractor, while in smaller populations, demographic noise overshadows the influence of selection effects. Thus, we establish that the dynamics of finite but sufficiently large populations can display hallmark characteristics of chaos when the underlying structure is inherently complex.

\subsection*{Data, materials and Software Availability}
Code base of the manuscript deposited on Zenodo (\url{https://doi.org/10.5281/zenodo.14660002}) \cite{ramirez:Codebase:2025}

\subsection*{Acknowledgements}
We thank Emil Mallmin, Silvia de Monte, Zachary P. Adams, Yuriy Pichugin, Henrik J. Jensen and Jürgen Jost for valuable feedback and discussions.
We also thank Ronald Kriemann for assistance with computer cluster usage at MPI-MIS.

\printbibliography
\end{refsection}

\beginsupplement
\fancyhead[L]{Supplementary material}

\twocolumn[{\textbf{\large Supplementary material: Chaos and noise in evolutionary game dynamics}
\tableofcontents \vspace{7mm}}]

\begin{refsection}

\section{Construction of the ACT interaction matrix}
\label{sec:construction_ACT}
After it was shown that ecological models of competing species could exhibit chaotic dynamics \cite{may:SIAMJAM:1975, smale:JMB:1976}, Arneodo, Coullet and Tresser (ACT) constructed a Shilnikov-type attractor for a Lokta-Volterra system with the following linear growth rates $\gamma_1=1.1$, $\gamma_2=-0.5$, $\gamma_3=1.75$, and interaction matrix
\begin{equation}
\label{eq:ACT_matrix}
\bordermatrix{ &  &  & \cr
                     & 0.5 & 0.5 & 0.1 \cr
                     & - 0.5 & -0.1 & 0.1 \cr
                     & 1.55 & 0.1 & 0.1 \cr}
\end{equation}

\begin{figure}[h]
      \centering
      \includegraphics[width=0.98\linewidth]{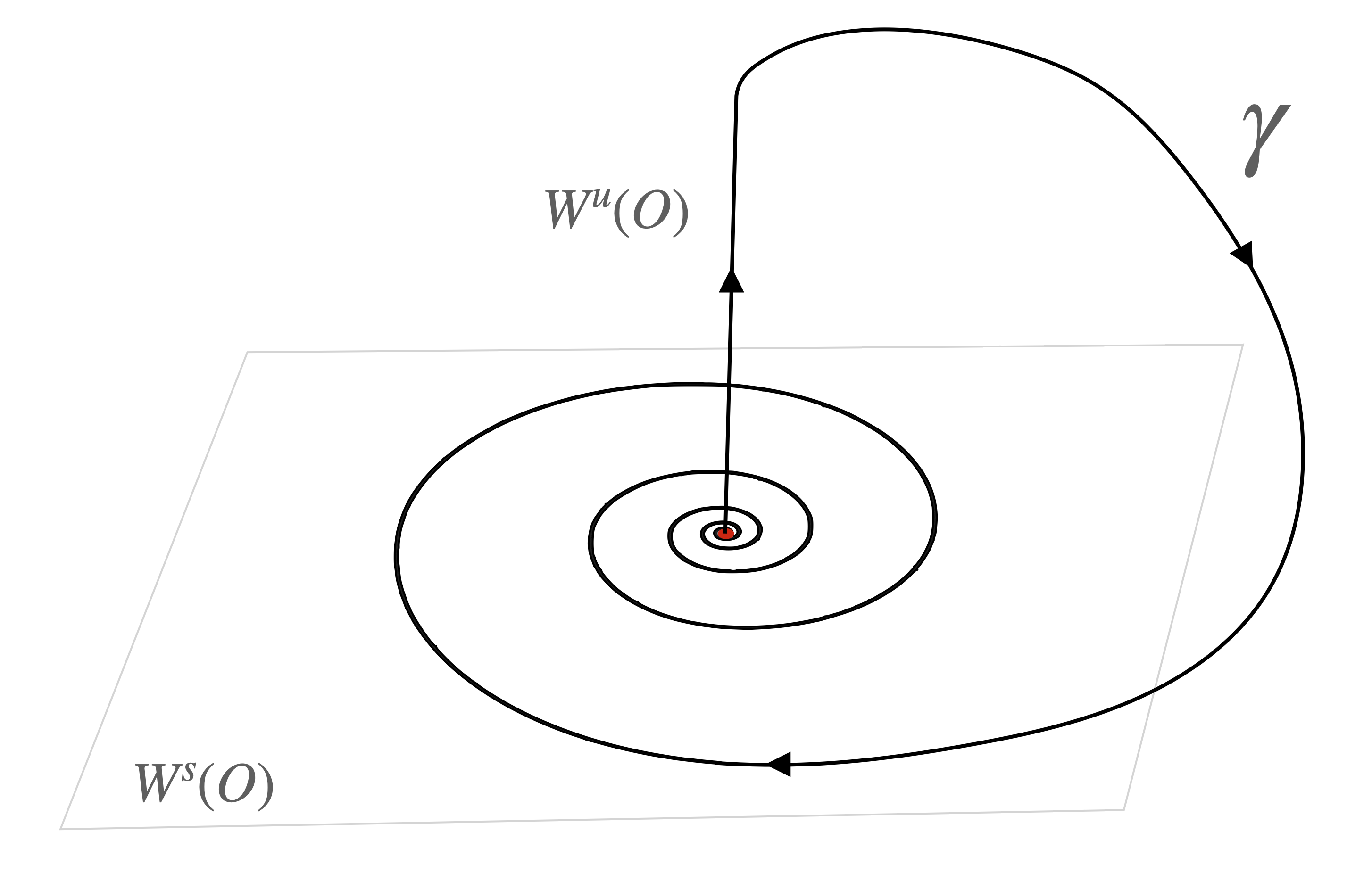}
      \caption{Shilnikov attractor in 3D phase space. The trajectories are drawn toward the saddle-focus equilibrium along its two-dimensional stable manifold, $W^s(O)$, but are repelled by its one-dimensional unstable manifold, $W^u(O)$. The Shilnikov attractor is created when the trajectories return to the equilibrium point through a homoclinic orbit, $\gamma$.}
      \label{fig:SI_spiral}
\end{figure}

\cite{arneodo:PLA:1980, arneodo:JMB:1982}. The construction is based on Shilnikov's theorem which explains how complex dynamics can emerge from a saddle-focus \cite{shilnikov:PUAS:1965}. Specifically, the theorem provides the conditions under which a dynamical system exhibits chaotic dynamics due to the presence of a homoclinic orbit around a saddle-focus (see Fig. \ref{fig:SI_spiral}). When these conditions are met, every neighbourhood of the orbit contains infinitely many unstable periodic solutions of saddle type. Moreover, the dynamics near the homoclinic orbit can be described by a Smale horseshoe structure, which represents the stretching and folding of trajectories that lead to chaotic motion \cite{guckenheimer:book:2013, hirsch:book:2013}.

In their paper, ACT present a heuristic method for constructing systems where Shilnikov's theorem applies. The construction centers on bifurcations and the trajectories around a saddle-focus fixed point, along with another fixed point that becomes a stable periodic orbit through a supercriticial Hopf bifurcation. Chaotic dynamics arise when the unstable manifold of the saddle-focus converges to the periodic orbit. When the attractor grows faster than the distance between the two points, the unstable manifold returns close enough to the saddle-focus to become part of its stable manifold generating a homoclinic orbit of Shilnikov's theorem. ACT point out that  this construction is not possible when all the entries of interaction matrix are positive because no focus can exist \cite{arneodo:PLA:1980}.

Interestingly, the phenomenon of Shilnikov chaos has been observed in a wide range of models, including those in chemistry, physics, economics and neuroscience. Based in these observations, it has been suggested that a universal mechanism may underlie the formation of Shilnikov or spiral chaos \cite{shilnikov:book:2001}.

\section{Numerical methods}
The dimensionality and nonlinearities required for chaotic dynamics to emerge in continuous dynamical systems make analytical methods impractical for such task. To solve the deterministic equation describing the pairwise comparison process numerically, we employ the Tsitouras 5/4 Runge-Kutta method of the \verb|OrdinaryDiffEq| Julia library \cite{rackauckas:JOSS:2017} via the \verb|DynamicalSystems| framework \cite{datseris:JOSS:2018}, as shown in the project's codebase repository \cite{ramirez:Codebase:2025}.

\section{Additional attractor visualisation}
\begin{figure}[h]
      \centering
      \includegraphics[width=0.80\linewidth]{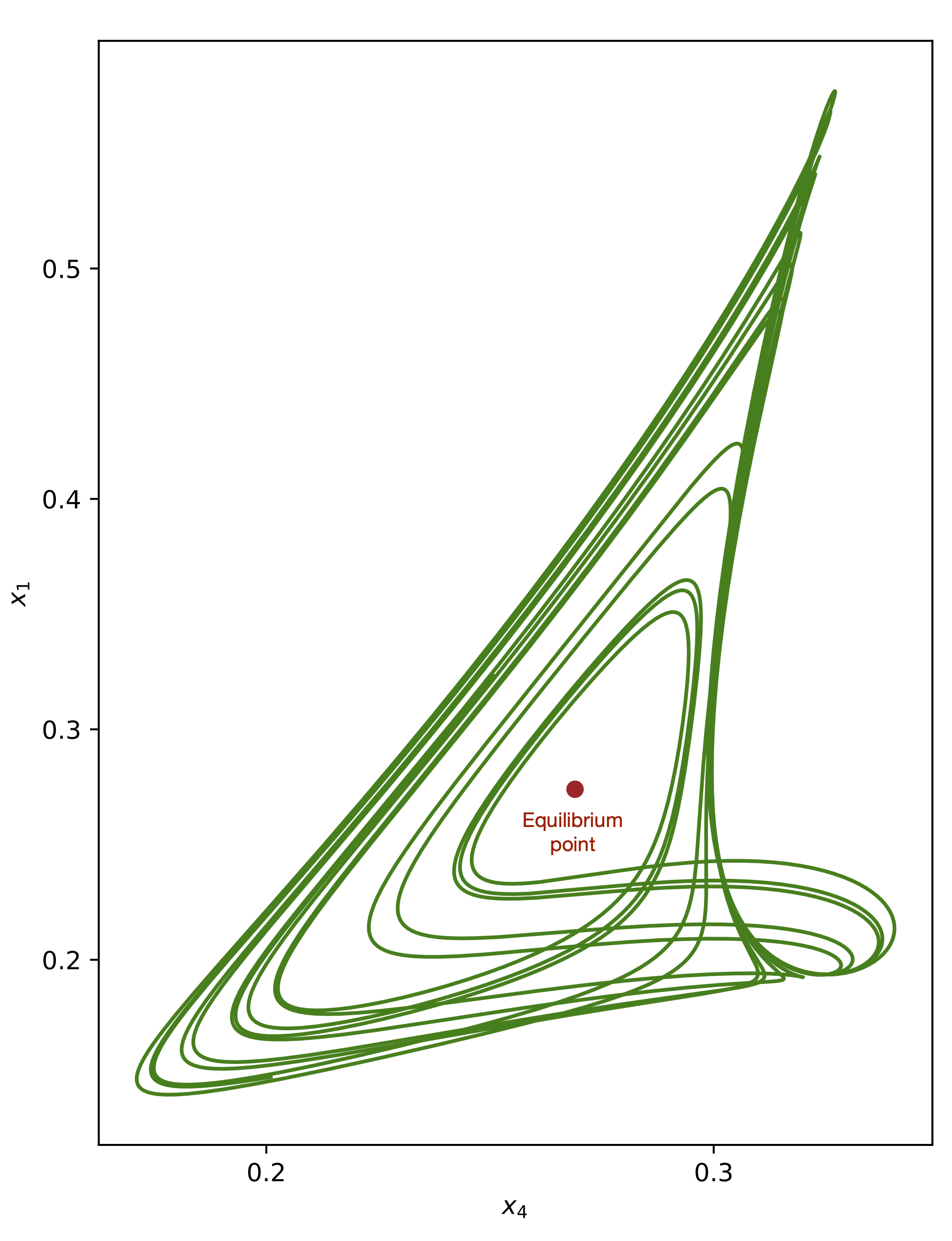}
      \caption{Chaotic attractor projected into the $x_1 x_4$ plane. The deterministic pairwise comparison process displays chaotic dynamics for the ACT-Skyrms payoff matrix under low selection intensity. For $\beta=0.1$, the trajectories orbit around the equilibrium point, $x^*$, following the structure of the Shilnikov attractor.}
      \label{fig:SI_planarview}
\end{figure}

Figure \ref{fig:SI_planarview} displays the chaotic attractor of the deterministic PCP projected into the $x_1 x_4$ plane. The trajectories start in $\vec{x}=(0.25,0.25,0.25,0.25)$ near the equilibrium point $x^* = (0.274, 0.215, 0.241, 0.269)$. If the initial conditions are near the boundary, the trajectories can reach extinction without entering the chaotic attractor's basin of attraction.

\section{Rescaling}
\label{sec:appendix_rescaling}

\subsection{Deterministic trajectories}
The deterministic pairwise comparison process exhibits a rescaling in time by the selection intensity parameter $\beta$, namely, the change in time is inversely proportional to the value of $\beta$. In practical terms, this means that for smaller $\beta$ values more time steps, $T$, are required to view the attractor. Thus, to obtain an equivalent observation of the dynamics for all $\beta$, we fix the number of effective time steps, $\tau$.

The number of effective time steps is defined as $\tau = T \beta$, where $\tau=1 \times 10^4$ is a constant estimated from the dynamics. Precisely, $\tau$ is the number of time steps required to fully visualise the attractor for $\beta=1.0$.

For smaller values of $\beta$, it is increasingly resource-intensive to study the dynamics because more time steps, $T$, are required to observe the attractor. To optimise the code, we scale the sampling time as $Dt=0.1/\beta$, such that trajectories with more time steps are sampled less often. It is important to note that the solver's relative and absolute tolerances are fixed at $1 \times 10^{-9}$ for all $\beta$ values. The relative tolerance specifies the number of digits to which the solution should be correct, while the absolute tolerance defines the threshold below which the adaptive solver treats small values as zero. Therefore, the sampling time does not affect the precision of the solutions.

\subsection{Stochastic trajectories}
The individual-based simulations that generate the stochastic trajectories are influenced by two factors: the population size, $N$, and the selection intensity, $\beta$. 

First, we consider the effect of the population size, $N$. In each time step, the population undergoes a single birth-death event, which means the composition of the population can only change $\Delta\vec{x}=1/N$ per time step. Consequently, as the population size increases, more time steps are required to cover the same length of the attractor because each step represents a smaller change in the population composition. 

Next, we consider the impact of the selection intensity, $\beta$. For smaller values of $\beta$, the stochastic trajectories also require more time steps, $T$, to visualise the attractor, particularly when the population size is large enough that the dynamics begin to approximate the deterministic description of the system.
This relationship means that the number of time steps needed scales according to $T = (N/\beta) \kappa$, where $\kappa \approx 5 \times 10^3$ is a constant estimated from the dynamics. For instance, with $\beta=5.0$ and $N=100,000$, the number of time steps required is $T=1\times 10^8$. These parameters represent the estimated upper limits to execute and quantify the stochastic simulations across multiple runs within a reasonable timeframe in our computer cluster. If one wishes to explore smaller $\beta$ values or larger population sizes $N$, more time steps will be necessary to fully capture the attractor generated by the trajectories.

There is an exception concerning the computation of the fractal dimension, which is particularly resource-intensive (see \ref{sec:appendix_FD}). For the fractal dimension to be measured across multiple runs within a reasonable amount of time, the limiting parameters are $\beta=5.0$ and $N=3,000$. In this scenario, the estimated rescaling is given by $T=N*C$, where $C\approx 170$. For this population size, the scaling due to $\beta$ is less pronounced, making it unnecessary to rescale according to $\beta$.

\section{Chaotic dynamics quantification}
In this section, we describe the tools and methods used to quantify the chaotic dynamics:

\subsection{Lyapunov spectrum}
The Lyapunov spectrum quantifies the evolution of infinitesimal perturbations in characteristic directions over time. To represent a $k$-dimensional volume surrounding the point $\vec{x}$, we use a matrix $Y$ consisting of $k$ perturbation vectors. The instantaneous rate of change for each column vector in $Y$ is described by the Jacobian, meaning the entire dynamics of $Y$ is determined by the linearised dynamics:
\begin{equation}
      \label{eq:linear_dyn}
      \dot{\vec{x}} = f(\vec{x}) \hspace{5mm} \dot{Y} = J_j(\vec{x})
\end{equation}

Starting with an arbitrary orthonormal matrix $Y_0$ as the initial perturbations around $\vec{x_0}$, we can solve \eqref{eq:linear_dyn} to obtain $Y(t)$. The goal is to quantify how the initial volume $Y_0$ changes over time to $Y(t)$ along characteristic dimensions. To achieve this, we perform a QR decomposition of $Y$ \cite{press:book:2007}, decomposing it into a column-orthogonal matrix $Q$ and an upper-triangular matrix $R$, such that $Y(t)=Q(t)R(t)$. The matrix $Q$ describes how $Y$ rotates over time, while $R$ describes how $Y$ changes in size along each of the $k$ dimensions. In particular, the diagonal entries of $R$ are used to obtain the Lyapunov spectrum, given that the exponents quantify the logarithmic change in size of $Y$
\begin{equation}
      \label{eq:lyapunov_exponents}
      \lambda_i = \lim_{t \to \infty} \frac{1}{t} \ln R_{ii}(t)
\end{equation}

To accurately compute $Y(t)$ over a long time period $t$, it is important to address potential numerical issues. These issues arise from the exponential growth or decay of matrix elements and the rapid convergence of vectors towards the leading expanding direction. To overcome these challenges, the simulation interval $[0,T]$ is divided into shorter intervals of size $\Delta t$. After each time interval, the solution vectors are renormalised and orthogonalised. 
Consequently, after $T$ steps the exponents can be approximated to 
\begin{equation}
      \label{eq:lyapunov_exponents_approx}
      \lambda_i = \frac{1}{T\Delta t}\sum_{i} \ln R_{ii} (t = i\Delta t)
\end{equation}
\cite{benettin:Meccanica:1980,geist:PTP:1990,datseris:book:2022}. There are various methods to compute the QR decomposition, in this case, we use the Householder reflection implemented in the Julia \verb|LinearAlgebra| library. The Householder reflection is a transformation that reflects a vector over a plane defined by the line bisecting the angle between the vector and a colinear reference vector. 
The complete Lyapunov spectrum computation is performed using the Julia \verb|ChaosTools| library.

\begin{figure}[h]
      \centering
      \includegraphics[width=0.98\linewidth]{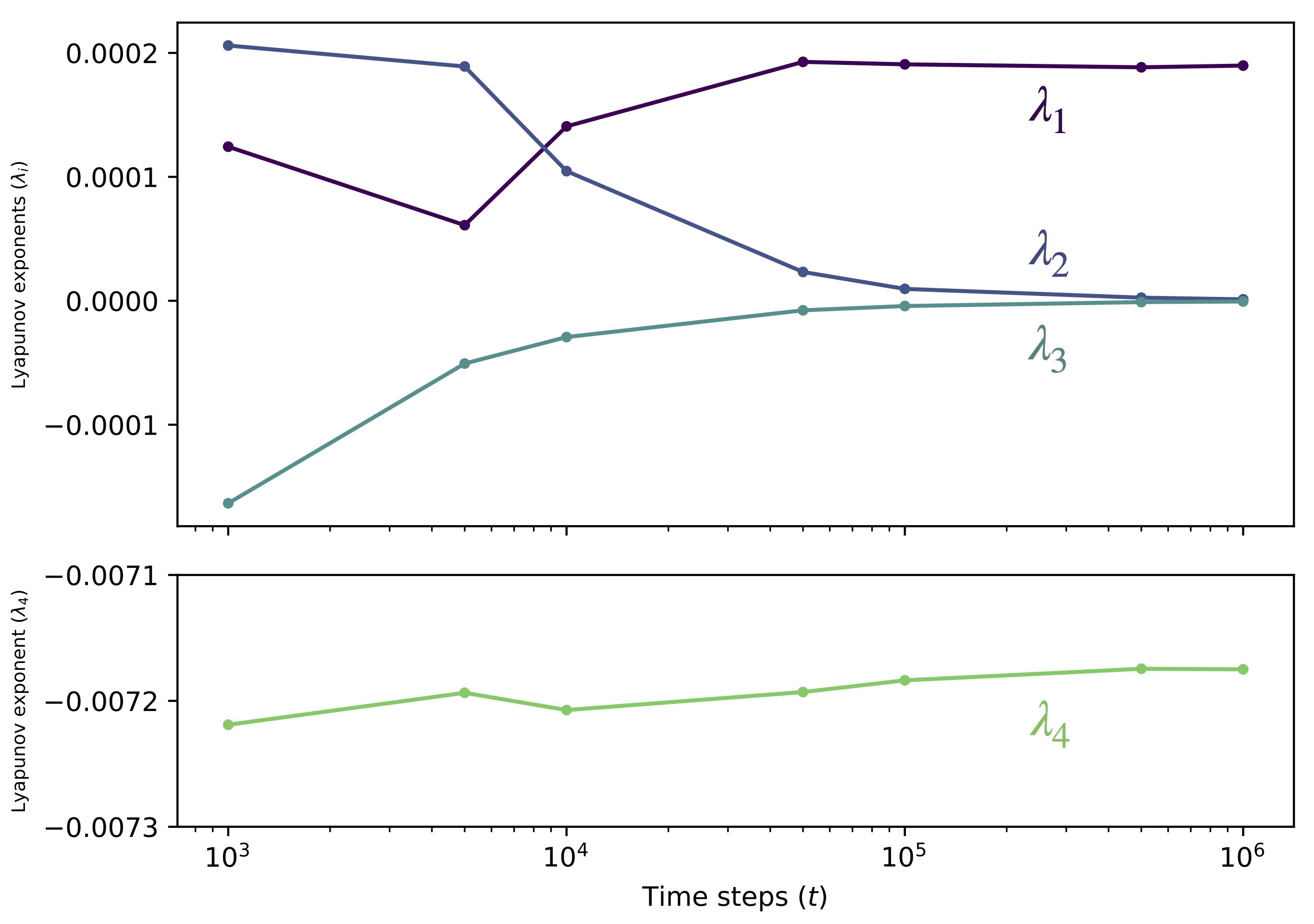}
      \caption{Convergence of Lyapunov exponents. The plot shows the number of time steps required for the Lyapunov exponents to converge for trajectories with $\beta=0.1$. This dynamics is representative of any $\beta$ values, as it is an inherent characteristic of the system. Overall, a minimum of approximately $T=1\times 10^5$ time steps is needed for the Lyapunov exponents to reach convergence.
      The Lyapunov exponents measures shown in the main text are based on $T = 5 \times 10^5$ time steps.}
      \label{fig:SI_LE}
\end{figure}


\subsection{Lempel-Ziv complexity}
The Lempel-Ziv complexity ($C_{LZ}$) quantifies the repetitiveness of a finite sequence, such as a string of text or binary sequence. It is measured by counting the number of distinct substrings or patterns encountered as the sequence is parsed from left to right. The quantification works by identifying the smallest unique substrings that have not been previously encountered in the sequence, thereby constructing a disctionary of patterns \cite{ziv:IEEEISIT:1978}. A higher $C_{LZ}$ indicates a more complex or less repetitive sequences, characterised by a greater number of unique patters, while a lower Lempel-Ziv complexity suggests more repetition or periodicity in the data set.

 To analyse the time series, we binarise the data by representing each data point as $1$ if it is above the time series' average, and as $0$ if it is below this threshold.

\subsection{Fourier spectrum}
\label{sec:appendix_fourier}
\begin{figure}[h]
      \centering
      \includegraphics[width=0.98 \linewidth]{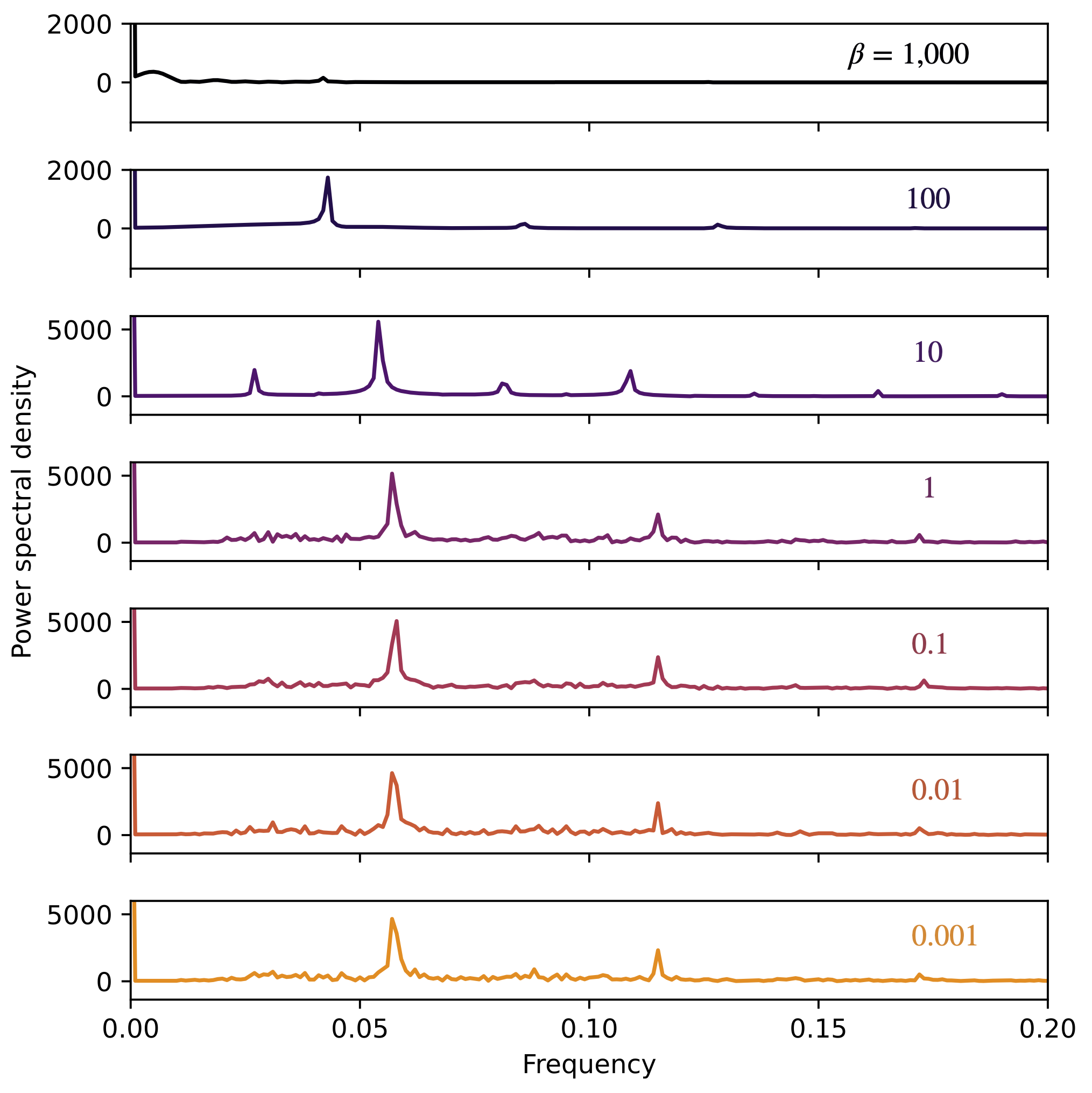}
      \caption{Power spectra of the deterministic PCP trajectories. For low selection strength, $\beta < \beta^*$ where $\beta^* \approx 7$, the system exhibits aperiodic dynamics characterised by a broadband spectrum between delta functions. For high selection, $\beta < \beta^*$, the spectral broadening disappears, indicating the emergence of periodic dynamics.}
     \label{fig:fourier_spectra}
\end{figure}

The Fourier transform is a method to identify aperiodic dynamics, which converts a function from its time domain into the frequency domain. In particular, we analyse the Fourier power spectra, as it represents the distribution of frequencies within a given trajectory. Figure SI.\ref{fig:fourier_spectra} shows the resulting spectra for different $\beta$ values.
For low selection intensity, the power spectra display a broadband spectrum between delta functions describing peaks at specific frequencies. In deterministic systems, this spectral broadening is a hallmark of chaos \cite{farmer:ANAS:1980}. Conversely, for larger ($\beta$) values, the broadening disappears, which is in agreement with the emergence of periodicity in this regime.

On a more technical level, the Fourier transform provides information about the frequencies present in the dynamical behaviour of a system. To analyse the frequency content, we perform a Fourier analysis using the Fast Fourier Transform (FFT), a well-known optimised algorithm that computes the discrete Fourier transform (DFT) of the input data. The DFT transforms the sequence ${x_n:=x_0,x_1,...,x_{N-1}}$ into ${X_k:=X_0,X_1,...,X_{N-1}}$, defined as

\begin{equation}
      \label{eq:DFT}
      X_k = \sum_{n = 0}^{N - 1} x_n e^{ - i2\pi kn /N} \hspace{5mm} k =0,..,N -1,
\end{equation}
where $\omega=e^{-i2 \pi /N}$ is the fundamental frequency. It is common practice to express the DFT as a transformation matrix, known as the DFT matrix, which can then be applied to the data set using matrix multiplication. Essentially, the DFT matrix is a $N\times N$ matrix expressed in terms of $\omega$. Thus, the DFT computation complexity is $\mathcal{O}(N^2)$. To make the process more efficient, the FFT factorises the DFT matrix into a product of sparse factors, which reduces the complexity to $\mathcal{O}(N log N)$ \cite{brunton:book:2022}. In this study, we use the \verb|FFTW| Julia package to compute the Fast Fourier Transform (FFT). This package provides bindings to FFTW, a C subroutine library designed for computing the discrete Fourier transform \cite{frigo:PIEEE:2005}. 

We analyse the Fourier power spectra, which plot the power spectral density - the squared magnitude of the frequency-domain function- against the corresponding frequency components. Figure \ref{fig:fourier_spectra} displays the Fourier spectra of variable $x_1$, which is representative frequency distribution of the other system's variables.

\subsection{Fractal dimension}
\label{sec:appendix_FD}

To calculate the fractal dimension ($\Delta^{c}$), we use the correlation dimension. The correlation dimension measures how tightly clustered are points in a set by counting the amount of neighbours within a given radius ($\varepsilon$). More precisely, it quantifies the probability that two randomnly chosen points are within $\varepsilon$ of each other. Here, $\varepsilon$ represents the radius of a hypersphere, within which points are considered neighbours \cite{grassberger:PRL:1983}. The correlation sum is defined as

\begin{equation}
      \scriptsize
      \label{eq:Correlation_sum}
      C(\varepsilon) = \frac{2}{(N - w)(N - w - 1)}\sum_{i =1}^{N - w - 1}\sum_{j = 1 + w + i}^{N}B( \| \vec{x_i} - \vec{x_j} \| < \varepsilon)
\end{equation}
where $\|\cdot\|$ represents a distance metric, in this case Euclidean, and $B=1$ if it's argument is true, $0$ otherwise. $N$ is the length of the data set and $w$ is the Theiler window, which allows the elimination of spurious correlations arising from dense time sampling \cite{theiler:PRA:1986}. To compute the correlation sum ($C$) for each radius value ($\varepsilon$), we use the box-assisted correlation sum, an optimised algorithm that encloses the data points into boxes to then calculate $C$ for each box, as well as for its neighbouring boxes \cite{theiler:PRA:1987}. We use the \verb|FractalDimensions| Julia package to measure the correlation sum.

\begin{figure}[h]
      \centering
    \includegraphics[width=1.0\linewidth]{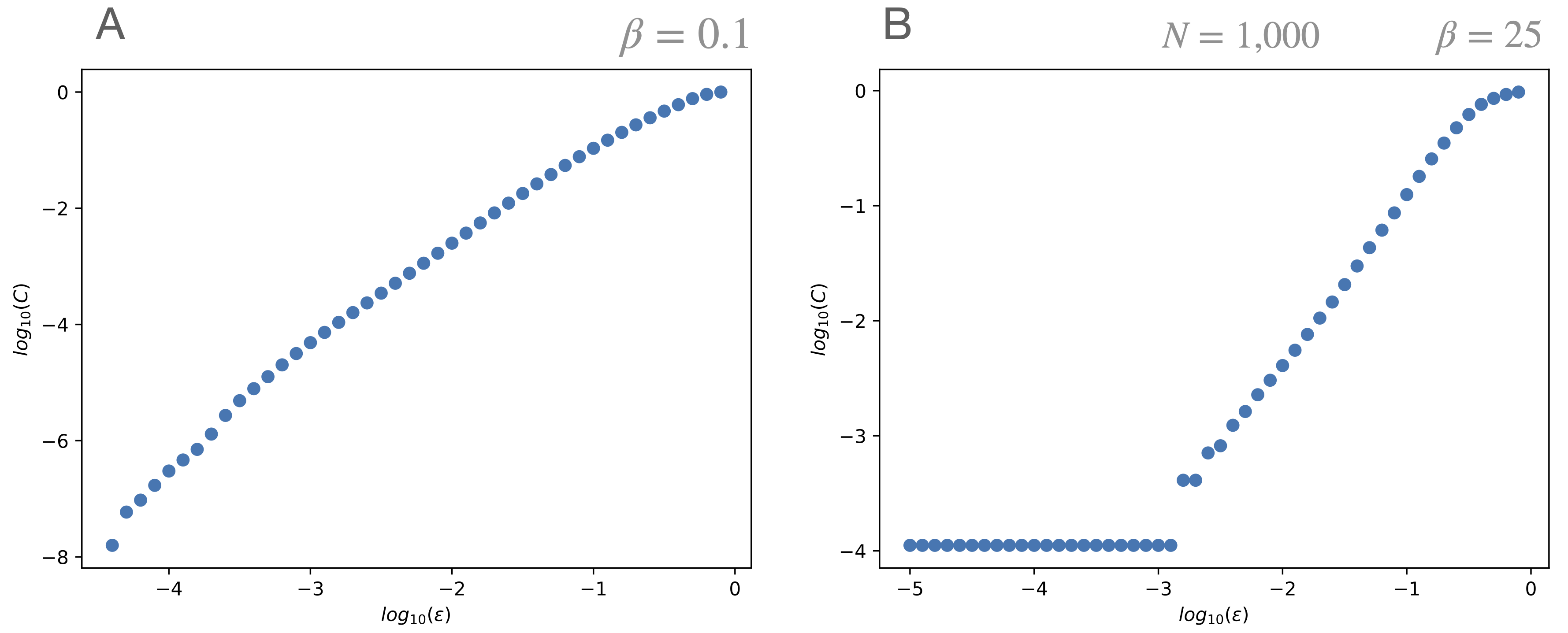}
      \caption{Fractal dimension estimates using the correlation sum ($C$). \textbf{A.} The deterministic trajectories display a long linear scaling region for $\log(C)$ versus $\log(\varepsilon)$ because the trajectories display fractality in a single spatial region. \textbf{B.} For stochastic trajectories, 
      the linear scaling region is present between the upper radius limit, determined by the effective radius ($R$), and the noise fine structure. At smaller scales, the stochastic trajectories' step size ($\delta$) leads to a jump at $\delta = 1/N \approx 10^{-3}$. For radii smaller than this value, there are no neighbouring points in the data set.}
      \label{fig:fractal_dim}
\end{figure}

From the correlation sum,  the scaling law 
\begin{equation}
\label{eq:ScalingLaw_correlationSum}
C \propto \varepsilon^{\Delta^{c}}
\end{equation}
follows.
Consequently, the correlation dimension is formulated as 
\cite{datseris:book:2022}
\begin{equation}
\label{eq:Correlation_dimension}
\Delta^c = \lim_{\varepsilon \to 0} \frac{\log(C)}{\log(\varepsilon)}
\end{equation}
In practice, the theoretical limit of $\varepsilon \to 0$ cannot be reached because data sets are finite. Hence, there is a minimum value of $\varepsilon$ for which the correlation sum ($C$) can be computed; below this value, $C$ becomes $0$. Similarly, there is an upper limit for the correlation sum. If $\varepsilon$ exceeds the attractor's effective radius ($R$), $C$ reaches $1$. Therefore, to accurately calculate $C$, it is necessary to choose an appropriate range of $\varepsilon$ values such that the plot of $\log(C)$ versus $\log(\varepsilon)$ shows a linear scaling region. The fractal dimension ($\Delta^{c}$) can then be determined from the slope of this linear region (see figure \ref{fig:fractal_dim}).

Figure \ref{fig:fractal_dim}.A shows the linear scaling region of deterministic trajectories. The plot displays a single linear region because the trajectories display fractality due the presence of a strange attractor.
Figure \ref{fig:fractal_dim}.B illustrates the linear scaling region of stochastic trajectories. Around $\log(\varepsilon) \approx -2.7$ ( or $\varepsilon \approx 0.002$), the fine structure of noise becomes apparent as a small plateau in the plot. For smaller values of $\varepsilon$, a jump occurs approaching $\log(\varepsilon) \approx -3$. This jump corresponds to the step size of the stochastic trajectories, $\Delta \vec{x} = 1/N$, which in this case is $10^{-3}$. Consequently, for radii smaller than $\Delta \vec{x}$, there are no neighbouring points in the data set.

\subsection{Attractor's standard deviation}
The attractor's standard deviation is calculated by computing the sample standard deviation (see \eqref{eq:std}) for each variable of the data set. The average of these four standard deviations is then used as an overall measure of the attractor's variability. We employ the standard deviation function from the \verb|Statistics| Julia package.

\begin{equation}
      \label{eq:std}
      \sigma = \sqrt{\tfrac{1}{N - 1}\sum_{i = 1}^{N}(x_i - \vec{x})^2}
\end{equation}

\subsection{Normalised fixation time}
The fixation time measures the number of time steps required for one of the types within the population to become extinct. Graphically, it is represented as the number of time steps the trajectories take to reach a boundary imposed by the normalisation condition.
However, since the number of time steps varies for each trajectory depending on the parameters $\beta$ and $N$, it is necessary to normalise the fixation time to ensure comparability across different scenarios. The normalised fixation time is thus defined as the number of time steps a trajectory takes to reach a boundary, divided by the total number of time steps determined by the scaling in appendix \ref{sec:appendix_rescaling}.

\section{Challenges related to the quantification of stochastic trajectories}
Methods that cannot distinguish the fine structure of noise from the self-similarity of the strange attractor fail to provide satisfactory measurements of the dynamics \cite{cencini:JPA:2013}. 
For instance, the Fourier spectrum displays fractality as a broad band in the frequency spectrum, but the broad bands of chaos and noise are intertwined \cite{bradley:Chaos:2015}. Although removing noise from the time series might help, we aim to directly analyse the stochastic trajectory without a tailored processing of the data. Thus, methods that require filtering out noise are not suitable for our problem.

Furthermore, methods that depend on data length are ineffective for quantifying the stochastic trajectories because demographic noise leads to fixation at different steps in the system's evolution, resulting in time series of different lengths. Consequently, a quantifier that is highly sensitive to data length cannot reliably characterise the stochastic trajectories accross different parameter values. For example, measures like the Lempel-Ziv complexity and related quantifiers are inadequate for this purpose.

Finally, methods that are sensitive to perturbations are not suitable to quantify the stochastic trajectories. The main quantifiers affected by this issue are the Lyapunov exponents, which measure a system's stability in response to infinitesimal perturbations. In our system, the perturbations are caused in each time step by demographic noise. As a result, estimates of the Lyapunov spectrum are unstable for the stochastic description, potentially leading to an unsuccessful quantification.
An alternative approach often used when analysing experimental data is to first embed the trajectories to reconstruct the underlying state space and then measure the desired quantifiers \cite{guckenheimer:Nature:1982,farmer:PRL:1987,sugihara:Nature:1990}. However, this nonparametric method is unnecessary in our case since we have complete knowledge of the underlying model and the state space of the trajectories. Thus, an embedding would not provide additional insight.

\printbibliography
\end{refsection}

\end{document}